# Search for power-efficient wide-range reversible resistance modulation of VO$_2$ single crystals

Bertina Fisher,[1,a)] Larisa Patlagan,[1] and Lior Kornblum[2]

[1]*Department of Physics, Technion – Israel Institute of Technology, Haifa 32000-03, Israel*

[2]*Andrew & Erna Viterbi Department of Electrical Engineering, Technion – Israel Institute of Technology, Haifa 32000-03, Israel*

The abrupt metal insulator transition in VO$_2$ is attracting considerable interest from both fundamental and applicative angles. We report on DC I-V characteristics measured on VO$_2$ single crystals in the two-probe configuration at several ambient temperatures below the insulator-metal transition. The insulator-mixed-metal-insulator transition is induced by Joule heating above ambient temperature in the range of negative differential resistivity (NDR). In this range the stability of V(I) is governed by the load resistance $R_L$. Steady state I(V) is obtained for $R_L > |dV/dI|_{max}$ in the NDR regime. For $R_L < |dV/dI|_{max}$ there is switching between initial and final steady states associated with peaks in the Joule power, that are higher the lower $R_L$ is. The peaks caused by steep switching are superfluous and damaging the samples. On the other hand, the large $R_L$ needed for steady state is the main power consumer in the circuit at high currents. The present work is motivated by the need to avoid damaging switching in the NDR regime while reducing the power consumption in the circuit. It is shown here that large resistance modulation can be obtained under steady state conditions with reduced power consumption by increasing the ambient temperature of the device above room temperature.

---

[a)] Author to whom correspondence should be addressed. Electronic mail: phr06bf@physics.technion.ac.il

Due to the outstanding insulator to metal transition (IMT) of VO$_2$ (T$_{IMT}$=340 K, resistance jump of up to five orders of magnitude in single crystals accompanied by large changes in its structural and optical properties) this material persists for decades as a candidate for many potential applications,[1] the most recent of which is for a new generation of electronics and electro-optics.[2–11] At the same time, recent studies continue to unveil fascinating mechanisms for these electronic features.[12–14] The insulator-metal (I-M) switching, whether induced thermally, optically or by an electric current is complex and induces a variety of mixed metal-insulator states and non-uniform stresses that produce mechanical degradation. Most of the investigations are focused on thin films which are more robust than VO$_2$ single crystals. Here we report on switching induced by an electric current in single crystals. Under an applied voltage at ambient temperature, the resistance drops due to Joule heating, the voltage reaches a maximum (V$_{max}$) and a current controlled negative differential resistivity (CC-NDR) regime sets on. The mixed insulator-metal (I-M) phase appears within the NDR regime. The stability of I(V) in this regime is governed by the magnitude of the load resistance (R$_L$) in series with the sample; I-V is stable for R$_L$ ≥ |dV/dI|$_{max}$ - the absolute value of the steepest slope of V(I) in the NDR regime.[15] For lower R$_L$, portions of the NDR regime are unstable and I(V) switches from the last steady state towards a steady state with minimal entropy production (minimal Joule heating).[16]

While in films the mixed state consists of metallic filaments embedded in the semiconductor,[17] in single crystals it consists of metallic and semiconducting domains with boundaries crossing the widths of the samples in favorable inclinations.[18] It was recently shown that in contrast to thin-film channels in Mott-FETs, single- crystalline nanowires have superior sensitivity for transport modulation, resulting in ten-fold higher resistance range; this is attributed to the coexisting I-M states in the crystalline channel.[19] VO$_2$ single crystals or free-standing, crystalline nano-wires have an additional, unique property: in the mixed M-I state: narrow semiconducting domains slide along a metallic background, in the positive sense of the electric current[20–23] being accompanied by relaxation oscillations.[22,24] This phenomenon has not yet found its use in applications but this may change in the future. The sliding domains are very sensitive to the integrity of the crystals; repeated I→M switching cycles or steep switching under low R$_L$ may cause the disappearance of sliding domains without having a significant effect on the ohmic properties of the samples;[18] upon repeated switching cycles, accumulated damage causes deterioration of all the properties of the samples. Upon switching between steady states in the NDR regime, the power P=IV passes through a maximum (P$_{max}$) that is steeper the lower R$_L$ is. Thus steep, damaging switching is prevented in VO$_2$ single crystals connected to large enough load resistances.

The threshold voltage V$_{max}$ at the onset of NDR regime may be estimated by equating the Joule power with Newton's law of cooling i.e., P=V$^2$/R=α∆T, where ∆T is the excess above ambient temperature T-T$_0$, and α is an effective coefficient. R is approximated by the activated resistance as R=R$_0$·exp(-∆T/T$_1$) where R$_0$ is the resistance at ambient temperature, and T$_1$ a fitting parameter.[25] This is allowed by the narrow temperature range between room temperature and T$_{IMT}$ (e.g. Fig. 1). For single crystals with activation energies of 0.4 ≤ E$_a$≤ 0.5 eV, this parameter is: 22 ≥ T$_1$ ≥ 17 K. These approximations lead to the following relation in the steady state (see Section A of Supplementary Material):

$$\frac{dV}{dI} = \frac{V}{I}\frac{(\alpha T_1 - IV)}{(\alpha T_1 + IV)} = \frac{V}{I}\frac{(\frac{T_1}{\Delta T}-1)}{(\frac{T_1}{\Delta T}+1)} \quad (1)$$

Eq. 1 shows that with increasing current (increasing self-heating) the differential resistance (dV/dI) decreases from V/I towards zero for $T_1 = \Delta T$ (independent of $\alpha^{25}$), at the onset of NDR, where $R(V_{max}) = R_0/e$ and $V_{max} = (R_0 e^{-1}\alpha T_1)^{1/2}$; the limit $T_1 \ll \Delta T$ for which dV/dR = -V/R cannot be reached since the onset of the mixed state is at finite $\Delta T = T_{IMT} - T_0$. For this model, $|dV/dI|_{max}$ in the NDR regime (that determines the minimal $R_L$ required for stability) as derived in Section A of the Supplementary Material is $0.0474 \cdot R_0$. Upon decreasing $R_0$ the maximal voltage on the sample (at the onset of NDR) is reduced and even more important, the load resistance (the main consumer of the applied voltage) can be reduced. The solution suggested in Ref. [18] for reducing $R_0$ while preserving the quality of $VO_2$ and its IMT would be to raise the ambient temperature around the functioning $VO_2$ device.

There may be another reason for increasing the ambient temperature of the $VO_2$ device close to $T_{IMT}$: when current is applied on the device at high temperature the I-M transition occurs at a low voltage that has no direct effect on the crystal except of heating. The high voltage needed to induce the transition in the more resistive device at low temperature may affect the lattice causing permanent alterations. The consequences of high voltages applied at low temperatures was shown to have dramatic effects in thin films.[13]

Finding optimal conditions for obtaining maximal resistance modulation under steady state conditions and minimal external voltage applied on single crystals is the main objective of this work. In addition, this work provides a test-bed of the simple model of "Joule heating in $VO_2$ single crystals".

Here we report on the investigation of the effects of the ambient temperatures and of $R_L$ (=101 kΩ, 11 kΩ and 1 kΩ) on the I-V characteristics of two single crystals of $VO_2$ (labeled D5(3), and D3(11), Table I). The results emphasize the striking difference between I-M switching under steady state ($R_L > |dV/dI|_{max}$) and non-steady state ($R_L < |dV/dI|_{max}$) conditions. The I-V characteristics of four more samples, D5(4), D3(16) and D3(17) were also measured at different ambient temperatures but only for $R_L > |dV/dI|_{max}$ in the NDR regime. I(V), R(V)=V/I and P(I)=IV(I) for samples D5(3) and D5(4) connected to different $R_L$ are presented in the main text, whereas those for D3(11), D3(16), D3(17) and D5(6) are shown in the Supplementary Material. Concluding results derived from all the traces in the main text and in the Supplementary Material are shown in the main text.

Crystal growth, sample preparation, contacts and I-V measurements are as described in the Supplementary Material of Ref. [18]. The semilog plots of R(1/T) for the six samples, measured by two contacts – four probes are shown in Fig. 1. The dimensions of the samples, their specific resistivity ρ at 300 K, the activation energy in the semiconducting state, the transition temperatures upon heating $T_{IMT}$ and upon cooling $T_{MIT}$ obtained from the traces in Fig. 1 are summarized in Table I. The resistance jumps are limited by the finite contact resistance in the metallic state (between ~ 7 Ω for D3(11) up to ~80 Ω for D3(16)). For all samples the transition is very steep and the hysteresis ranges between 1 K for D5(3) to 4 K for D5(6) – an indication of their high crystalline quality.

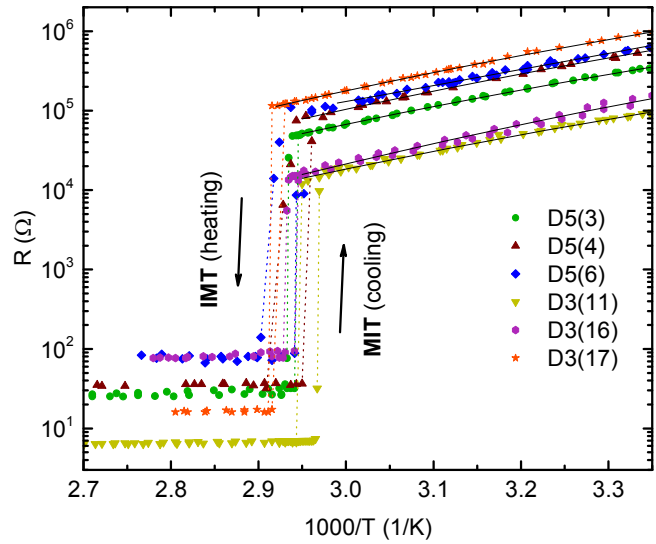

**Figure 1.** Two-contact resistance: R versus 1000/T for crystals D5(3), D5(4), D5(6), D3(11), D3(16) and D3(17).

**Table I.** Properties of the samples used in this work. Length (L), cross-section area (A), resistivity ($\rho$) at 300 K, activation energy ($E_a$), transition temperature upon heating ($T_{IMT}$) and upon cooling ($T_{MIT}$), extracted from Fig. 1. $V_{max}/R_0^{0.5}$ and $R^2$ represent the slopes in Fig. 5a and their fitting quality, respectively.

| Sample | L(cm) | a(cm) | b(cm) | $\rho$(300K) ($\Omega$cm) | $E_a$(eV) | $T_{IMT}$ (K) | $T_{MIT}$ (K) | Surface area (cm$^2$) | $V_{max}/R_0^{1/2}$ (W$^{1/2}$) | $R^2$ |
|---|---|---|---|---|---|---|---|---|---|---|
| D5(3) | 0.15 | 0.00736 | 0.00485 | 85 | 0.412 | 340.5 | 339.5 | 0.00366 | 0.0573 | 0.963 |
| D3(11) | 0.15 | 0.0210 | 0.0160 | 199 | 0.409 | 339 | 337 | 0.0111 | 0.0928 | 0.924 |
| D5(4) | 0.15 | 0.0144 | 0.0049 | 240 | 0.414 | 341 | 338 | 0.0058 | 0.0562 | 0.905 |
| D3(16) | 0.42 | 0.0260 | 0.0246 | 194 | 0.476 | 341 | 339.5 | 0.0425 | 0.1202 | 0.922 |
| D3(17) | 0.43 | 0.0088 | 0.0044 | 242 | 0.429 | 343 | 341 | 0.01136 | 0.0926 | 0.803 |
| D5(6) | 0.15 | 0.0077 | 0.0060 | 177 | 0.400 | 342 | 338 | 0.00411 | 0.0553 | 0.990 |

The I-V characteristics of D5(3) at 295 K, 310 K and 320 K, R(V)=V/I and P(I)=IV calculated from the I(V) data are shown in the first row of frames of Fig. 2 for $R_L$=101 k$\Omega$, in the second row for $R_L$=11 k$\Omega$ and in the third for $R_L$=1 k$\Omega$ (1 k$\Omega$ is the resistor used for the current measurements). In the first row, for the largest $R_L$, the I-V characteristics show the onset of CC-NDR at $V_{max}$ that decreases as the temperature is increased. Small voltage drops at a current which is independent of $T_0$ towards a reversible I(V) range mark the onset of the mixed state with the appearance of the first metallic domain [e.g. at ~20 V for 295 K, panel (a)]. The range of currents in (a) is limited by the large $R_L$, that carries most of the finite source voltage of 200 V. The hysteresis is narrow and the backward transitions occur at slightly lower currents. R(V) calculated from I(V) and |dV/dI|(V) from fitted cubic or higher order polynomials to V(I) in the NDR regime are shown in frame (b); the latter are shown below the R(V) curves. For all three ambient temperatures |dV/dI|$_{max}$<101 k$\Omega$ but >11 k$\Omega$. Thus the data in the NDR regime with $R_L$=101 k$\Omega$ were obtained under stable conditions while all others, under unstable conditions. Frame (c) shows the increase of P=IV with increasing current (increasing temperature) followed by a small but steep decrease as the first metallic domain appears in the sample, at a 13~15 mA range for all three temperatures. This is consistent with the decrease in infrared emission of metallic VO$_2$ relative to semiconducting VO$_2$.[26,27]

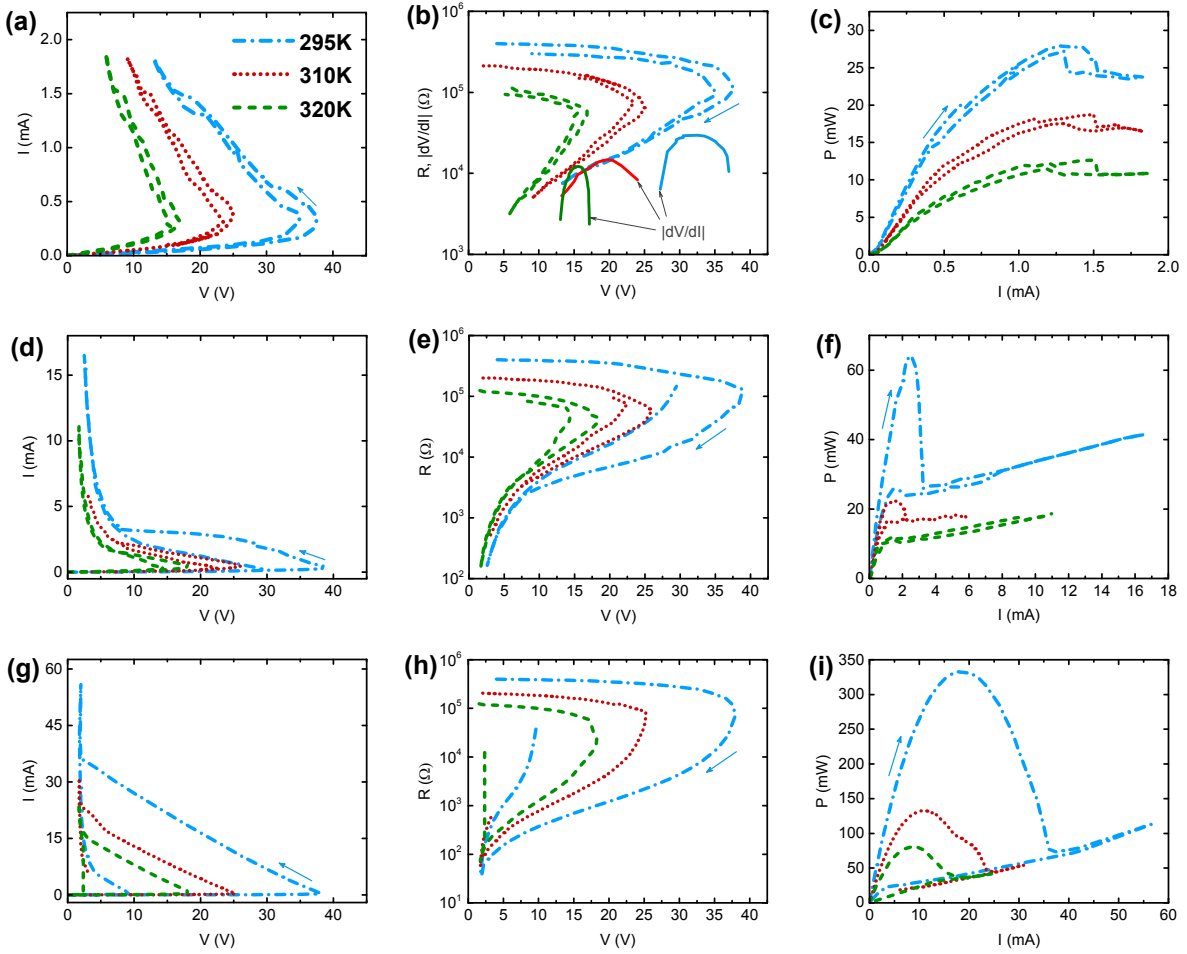

**Figure 2.** I(V), R(V)(=V/I) and P(V)(=VI) for sample D5(3) at temperatures 295, 310 and 320 K [see legend on frame (a)] with $R_L$=101 kΩ [(a) –(c)], $R_L$=11kΩ [(d) –(f)] and $R_L$=1 kΩ [(g)-(i)]. |dV/dI|(V) in the NDR regime is shown in (b) below R(V). Arrows indicate the scan direction.

The range of currents in Fig. 2(d) is one order of magnitude larger for $R_L$=11 kΩ compared to 101 kΩ. Switching occurs along a broken line, that is steeper close to $V_{max}$ and almost flat before the onset of the reversible (stable) regime. A straight line of slope ~1/$R_L$ connects the point at $V_{max}$ with that at the onset of the reversible state. I(V), R(V) and P(V) exhibit large hystereses between forward and backward switching. The lowest resistance reached upon increasing current (Fig. 2(e)) is still at a safe distance above that of the contact resistance (<0.1 kΩ, Fig. 1). $P_{max}$ for $R_L$=11 kΩ (Fig. 2(f)) is more than twice larger than for $R_L$=101 kΩ at 295K; interestingly, $P_{min}$ is close to that for the large $R_L$. The drop of $P_{max}$ with increasing temperature is dramatic. In the reversible (stable) regime, P increases linearly with I. This was quite surprising at the time when it was naively assumed that for fixed temperature ($T_{IMT}$) P should either be constant or decrease with current due to the increasing fraction of the metallic phase in the mixed state.[26] It was previously shown that the increasing P(I) is correlated to the dynamic phenomena of creation and annihilation of moving M-I domain boundaries. The slope dp/dj =L$^{-1}$dP/dI≈ 8 V/cm for this sample (p-power density, j-current density and L-sample's length) falls within the range of such phenomena observed in Ref. [18].

The range of currents for $R_L$=1 kΩ was intentionally limited to only a factor of ~3 relative to the previous case, so that the lowest resistance would not reach the contact resistance. Switching in Fig. 2(g) is along a straight line of slope ~1/$R_L$ from the onset of NDR to the reversible (stable) range of currents. The hystereses in all graphs of the three frames are huge. At 295 K, $P_{max}$ for 1 kΩ is larger than $P_{max}$ for 101 kΩ by more than an order of magnitude (Fig. 2(i)).

The effect of $R_L$ on $P_{max}$ is revealed in Figs. 3(a) for sample D5(3) and in 3(b) for D3(11) (Fig. S3, Supplementary Material). Semilog $P_{max}$($T_0$) representation was chosen in order to provide equal weights to the data that are spread over almost two orders of magnitude.

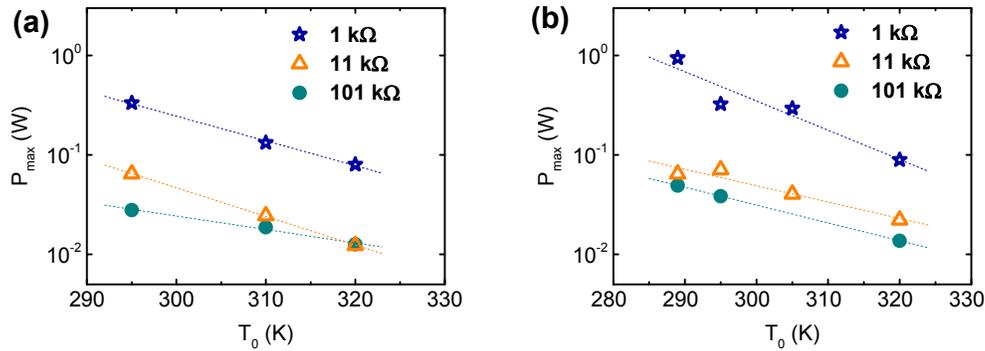

**Figure 3.** The peak power $P_{max}$ at switching versus ambient temperature $T_0$ for crystals D5(3) (a) and D3(11) (b) connected to the different load resistances $R_L$.

In Fig. 3(a) the fitted lines for $R_L$=1 kΩ and $R_L$=11 kΩ have relatively higher slopes and below them lies the line for $R_L$=101 kΩ with a much lower slope. In Fig. 3(b) the line at the top (for $R_L$=1 kΩ) has a high slope while the two at the bottom (for $R_L$=11 kΩ and $R_L$=101 kΩ) have relatively lower slopes. The data on the lines with high slopes were obtained under non-steady state conditions (i.e. $R_L$ < |dV/dI|$_{max}$ for the respective characteristic). With one exception, the data on the lines with low slopes were obtained under steady state conditions. The exception is the datum point for D3(11) at 289 K (see Fig. S3 in the Supplementary Material). It represents the I-V characteristic that deviates from the "well behaved" form in Fig. S3(f) and shows that this characteristic is for a non-steady state. In fact, a line for $R_L$=11 kΩ fitted to the three data points for the higher temperatures would be parallel to the line for $R_L$=101 kΩ. According to our simple model $P_{max}$ ∝ ($T_{IMT}$-$T_0$); for narrow ranges the exponential function is close to linear. Therefore, the overall behavior in Fig. 3 agrees with our proposed model where P∝ΔT in steady state.

I-V measurements on sample D5(4) (Fig. 4) were devoted to finding the optimal, steady state conditions for insulator - mixed metal-insulator transition induced by current, that is, maximal resistance modulation for minimal applied voltage. Steady state implies removal of the damaging bell shaped P(I) in the NDR regime (e.g. Fig. 2(f),(i)).

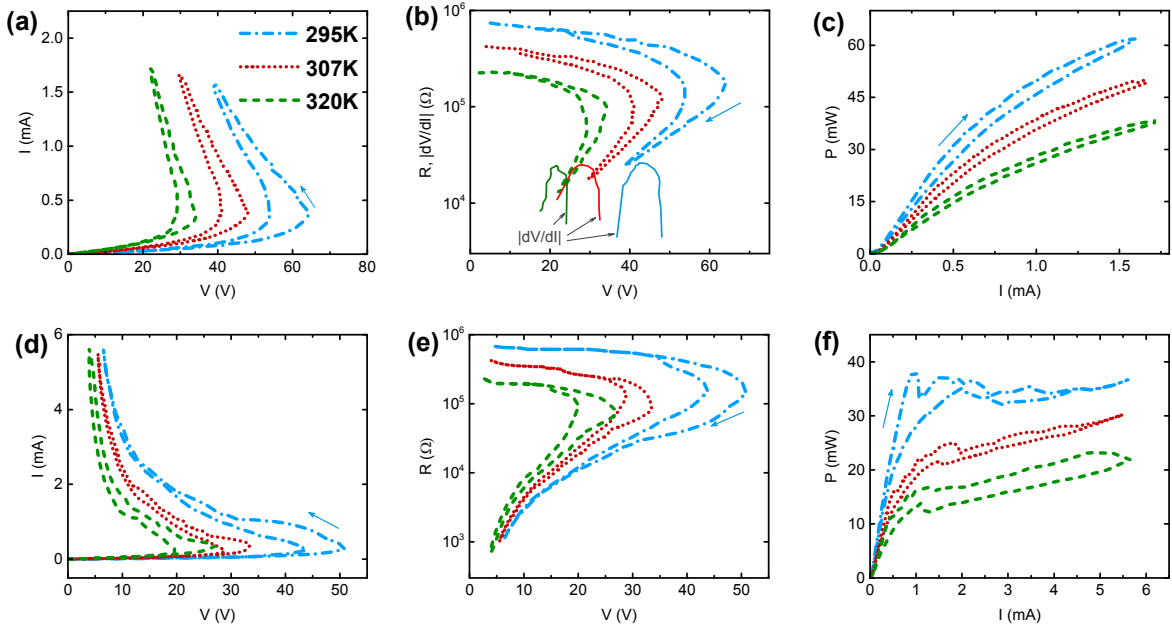

**Figure 4.** I(V), R(V)(=V/I) and P(V)(=VI) for sample D5(4) at temperatures 295 K, 307 K and 320 K [see legend in panel (a)] with $R_L$=101 kΩ [(a) –(c)] and $R_L$=34kΩ [(d) –(f)]. |dV/dI|(V) in the NDR regime is shown in (b) below R(V). Arrows indicate the scan direction.

The I-V characteristics, R(V) and P(I) for $R_L$=101 kΩ are shown in Fig. 4((a)-(c)) for sample D5(4). The traces of |dV/dI|(V) for this large $R_L$ are shown below the R(V) traces in Fig. 4(b). They show that |dV/dI|$_{max}$ = 26.1 kΩ, 25.0 kΩ and 24.6 kΩ for $T_0$=295 K, 307 K and 320 K, respectively. These traces for $R_L$=34 kΩ (out of caution, not the lowest possible $R_L$) are shown in Figs. 4((d)-(f)). The range of currents was increased by a factor of three and that of the resistance modulation by one order of magnitude. The lowest resistance in 4(e) is more than one order of magnitude higher than the contact resistance. The results in Fig. 4(f) are in strong contrast with those shown in Figs. 2(f) and 2(i) for sample D5(3) (and those of Fig. S3 (i) for sample D3(11)). For all three temperatures P(I) reaches shallow maxima followed by shallow dips associated with the appearance of metallic domains. After a range where P(I) shows a rather noisy behavior, reversible ranges are reached. The widest range of currents over which P(I) is reversible is obtained for $T_0$ = 320 K. The slope of this trace in the reversible regime is dP/dI ≈1.5 V (comparable to that in Fig. 2(f)) and corresponds to dp/dj= 10 V/cm. The low $V_{max}$ and the wide reversible regime emphasize the advantage of working at $T_0$=320 K. In other words, at low resistance it is distinctly adventitious to perform current-induced switching of $VO_2$ crystals close to their IMT temperature. This can be intuitively be expressed as switching in the conditions where less driving force is necessary, which promotes a 'clean' transition that results in reversible, damage-free behavior.

In Figure 5 we compare results obtained for the six samples by which the validity of the simple model of "Joule heating" is examined.

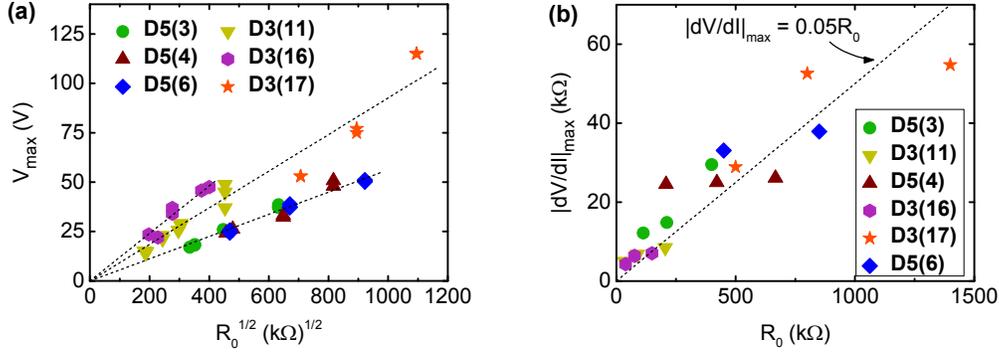

**Figure 5.** (a) $V_{max}$ versus $R_0^{1/2}$ for crystals D5(3), D3(11), D5(4), D5(6), D3(16) and D3(17). (b) $|dV/dI|_{max}$ versus $R_0$.

In Fig. 5(a) we plotted $V_{max}$ versus $R_0^{1/2}$ according to the model's predictions. In principle, for a given sample $V_{max}$ should depend only on temperature and not on the load resistance. This holds for all but sample D3(11) at 295 K. The slopes of the lines and the respective quality of fit are summarized in the last two columns of in Table I. The values of $R^2$ (> 0.9) for all but that for sample D3(17) ( = 0.8) indicate fairly good agreement between the experimental results and predictions. The fitted lines for D3(11) and D3(17) almost coincide and so do the three fitted lines for D5(3), D5(4) and D5(6). The surface areas of the first pair of samples is incidentally identical (Table I), those for the other three are close but a small uncertainty in their small dimensions could account for the difference between their areas. The correlation between $V_{max}^2/R_0$ and surface areas found for these five samples stresses the fact that the heat dissipation is almost exclusively from the surface of the samples, and therefore heat conduction through the contacts is insignificant.

The $|dV/dI|_{max}$ data versus $R_0$ obtained from the V(I) traces for the six samples in the stable states of NDR are plotted in Fig. 5(b). The data points are best fit by the line $|dV/dI|_{max}=0.05\,R_0$. The model predicts $|dV/dI|_{max}=0.0474\,R_0$ for NDR in the insulating state. The onset of the mixed M-I state changes the slope of dV/dI (V) from that derived for the insulating state. For $T_0$= 320 K (with $T_1$ around 20 K) the onset of NDR (at $T_0+T_1$) and of the mixed state at $T_{IMT}$ are very close. Therefore, to warrant steady state $|dV/dI|_{max}$ should be determined experimentally for each device at each ambient temperature. In a more general case, for applied AC, the load resistance $R_L$ should be replaced by the load impedance, $Z_L$.

In conclusion, our I-V measurements show the advantage of increasing the ambient temperature around single-crystalline devices for avoiding steep switching in the NDR regime while reducing power consumption. The results indicate that the simple "Joule heating model", though not rigorous, is very useful for orientation within the various regimes of the I-V characteristics. Large deviations of the measured $|dV/dI|_{max}$, calculated from the model (derived for the insulating regime) are caused by the appearance of the mixed I-M state within the NDR regime.

**SUPPLEMENTARY MATERIAL**

See supplementary material for the derivation of the model and Eq. (1), and for the full electrical analysis of samples D5(6), D3(11), D3(16) and D3(17).

**ACKNOWLEDGMENTS**

We thank Dr. George M. Reisner from the Physics Department of the Technion for reading the manuscript and for insightful discussion and comments. L.K. is a Chanin Fellow.

# Search for power-efficient wide-range reversible resistance modulation of $VO_2$ single crystals.

Bertina Fisher,[1,a)] Larisa Patlagan,[1] and Lior Kornblum[2]

[1]*Department of Physics, Technion – Israel Institute of Technology, Haifa 32000-03, Israel*

[2]*Andrew & Erna Viterbi Department of Electrical Engineering, Technion – Israel Institute of Technology, Haifa 32000-03, Israel*

**Supplementary Material**

### A. The simple model of Joule heating in $VO_2$ single crystals.

This model is based on two approximations:

1. The activated resistance in the semiconducting regime is approximated by an exponential decay of the resistance from $R_0=R(T_0)$, where $T_0$ is the ambient temperature to $R(T_{IMT})$:

$$R(T) = R_0 \exp\left[\frac{E_a}{kT} - \frac{E_a}{kT_0}\right] = R_0 \exp\left[\frac{E_a(T_0-T)}{kTT_0}\right] \approx R_0 \exp\left(\frac{-\Delta T}{T_1}\right)$$

where $\Delta T = T - T_0$ and the parameter $T_1 = kT_0 \langle T \rangle / E_a$. This approximation is valid for narrow ranges of temperatures relative to the average temperature $\langle T \rangle$. $T_1(E_a)$ is shown in Fig. A1 for two extreme ambient temperatures employed in most of this work, $T_0 = 295$ K and $T_0 = 320$ K. Since the maximal temperature is 340 K, $\langle T \rangle = 317.5$ K in the first case, and 330 K, in the second.

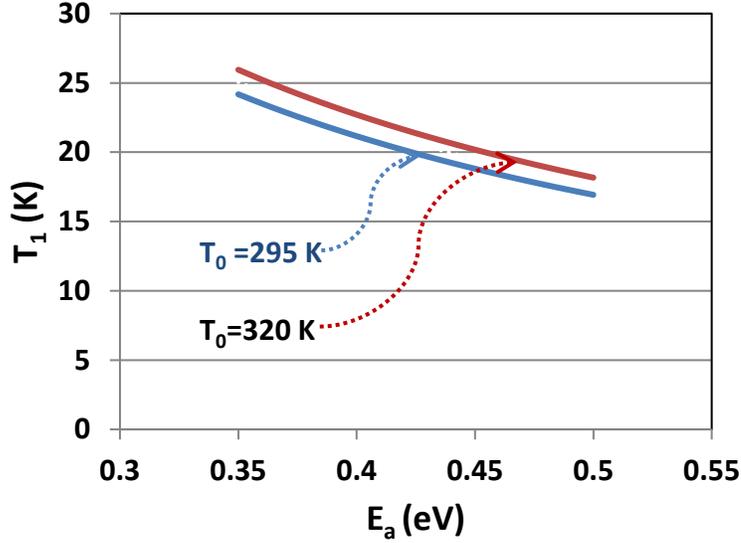

**Figure S1.** The parameter $T_1$ versus activation energy $E_a$ for two ambient temperatures $T_0$ employed in this work.

The Arrhenius law and the exponential dependence fit equally well the experimental R(T) of $VO_2$ single crystals measured between room temperature and 340 K.

2. The Joule power is equated with Newton's cooling law: $P=IV = \alpha \Delta T$ where $\alpha$ is a constant. This is a first approximation for various cooling processes valid for $\Delta T/T \ll 1$.

Using these approximations we obtain for fixed ambient temperature $\Delta T = T_1 \ln(R_0 I/V)$. The I-V characteristic is given by the relation:

$$IV = \alpha T_1 \ln\left(\frac{IR_0}{V}\right) \quad (1A)$$

From 1A we obtain the differential resistance $\frac{dV}{dI}$:

$$\frac{dV}{dI} = \frac{V(\alpha T_1 - IV)}{I(\alpha T_1 + IV)} \quad (2A)$$

At the onset of CC-NDR, $\frac{dV}{dI} = 0$. This yields $\Delta T = T_1$ and from 1A: $\frac{V}{I} = R_0 e^{-1}$ (independent on α) and $V_{max} = \sqrt{R_o \alpha T_1 e^{-1}}$.

In terms of normalized variables: $v=V/\sqrt{R_0\alpha T_1}$, $i=I\sqrt{R_0/\alpha T_1}$, $p=IV/\alpha T_1 = \Delta T/T_1$ and $v/i=V/(IR_0)=e^{-p}$. Eqs. (1A) and 2(A) become:

$$iv=\ln\left(\frac{i}{v}\right) \quad (1'A)$$

$$\frac{dv}{di}=\frac{v}{i}\frac{(1-iv)}{(1+iv)} \quad (2'A)$$

The $i-v$ characteristic, $\frac{v}{i}$ and $\frac{dv}{di}$ versus $v$ are shown in Fig. 2A.

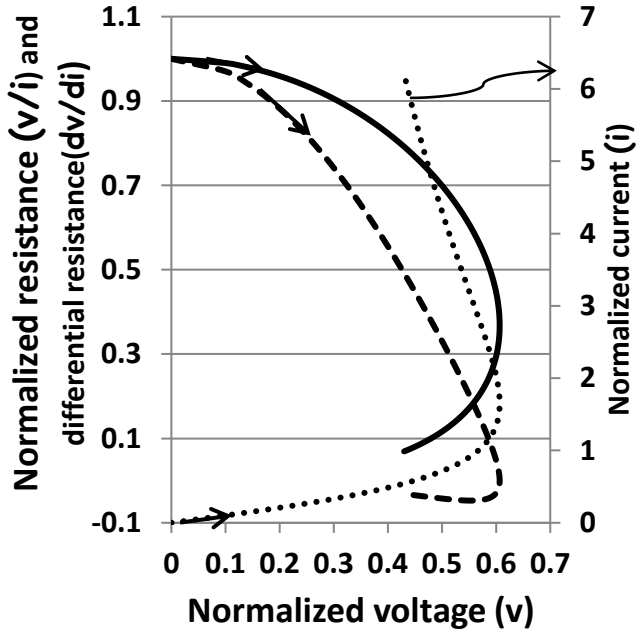

**Figure S2.** $\frac{v}{i}$ versus $v$ (solid line), $\frac{dv}{di}$ versus $v$ (dashed line) and $i$ versus $v$ (dotted line).

The condition for stability in the CC-NDR regime is $R_L > |\frac{dV}{dI}|_{max}$ obtained when $\frac{d^2V}{dI^2}=0$ (i.e. when $\frac{d^2v}{di^2}=0$).

$$\frac{d^2v}{di^2}=\frac{2v^2}{i}\left(\frac{3-(vi)^2}{(1+vi)^3}\right)=0 \quad (3A)$$

Eq. (3A) yields $p = vi = \sqrt{3}$, thus $\frac{v}{i}=e^{-\sqrt{3}}$ and $\frac{dv}{di}=e^{-\sqrt{3}}\frac{1-\sqrt{3}}{1+\sqrt{3}}=-0.004741$; $|\frac{dV}{dI}|_{max}=0.004741\,R_0$.

## B. Supplementary Experimental results.

In the main text of the manuscript we reported on the investigation of the effects of the ambient temperatures and of $R_L$ (=101 k$\Omega$, 11 k$\Omega$ and 1 k$\Omega$) on the I-V characteristics of a single crystals of $VO_2$ (labeled D5(3)), Table I. We reported also on the investigation of the effects of the ambient temperatures a single crystal (labeled D5(4), Table 1) but only for $R_L < |dV/dI|_{max}$. The results emphasize the striking difference between I-M switching under steady state ($R_L > |dV/dI|_{max}$) in the NDR regime and non-steady state ($R_L < |dV/dI|_{max}$) conditions. The results of the investigation of the effect of temperature on the I-V characteristics of four more samples, are presented here, the first (D3(11) with $R_L$ =101 k$\Omega$, 11 k$\Omega$ and 1 k$\Omega$ and the later three [D3(16), D3(17) and D5(6)] with $R_L > |dV/dI|_{max}$ in the NDR regime,

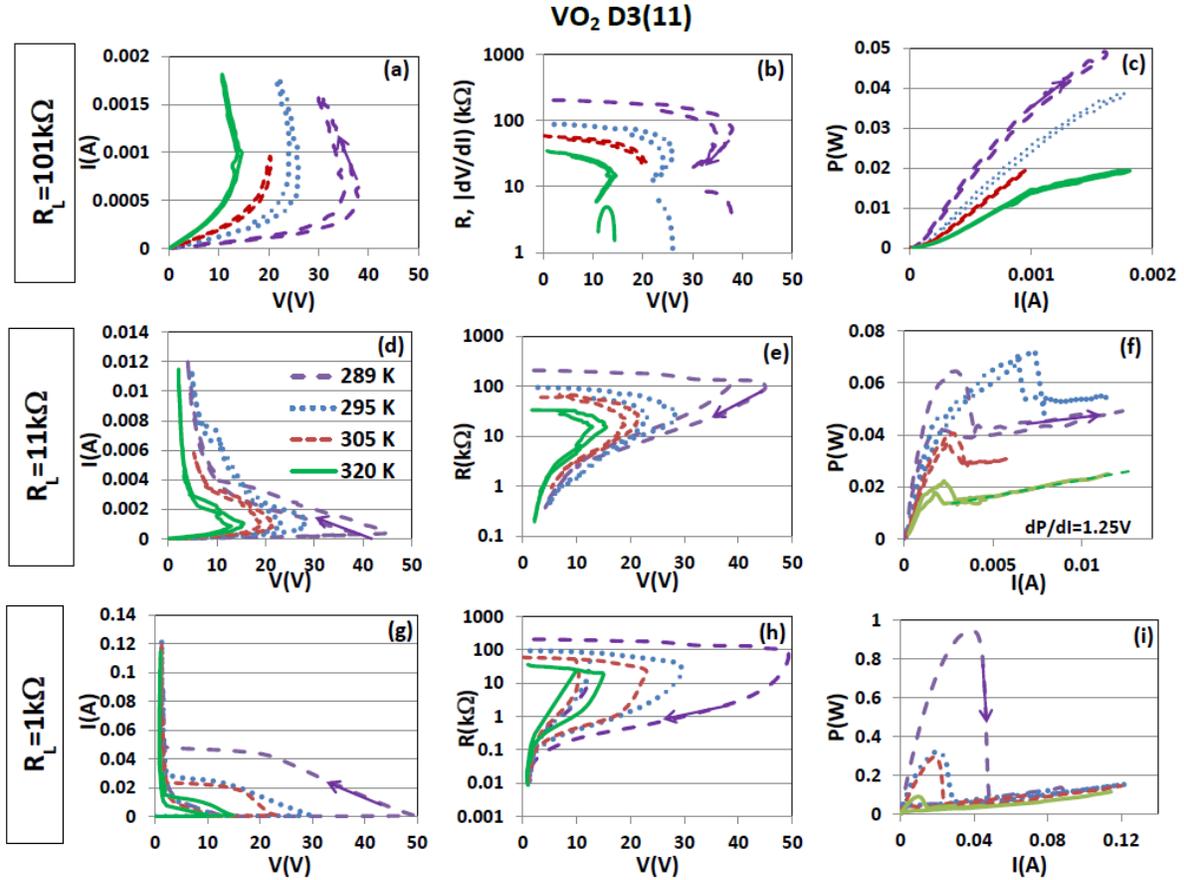

**Figure S3.** I(V), R(V)(=V/I) and P(V)(=VI) for sample D3(11) at temperatures 289, 295, 305 and 320 K (see legend on frame (d)) with $R_L$=101 k$\Omega$ ((a) –(c)), $R_L$=11k$\Omega$ ((d) –(f)) and $R_L$=1k$\Omega$ ((g)-(i)). $|dV/dI|$(V) in the NDR regime is shown in (b) below R(V).

The experimental results for D3(11) (I(V), R(V), $|dV/dI|$, P(I) ) at various temperatures and $R_L$ are presented in the same format as for D5(3) and therefore the main similarities and differences between the data for the two samples will be shown. Measurements on D3(11) were extended to a lower temperature, $T_0$= 289 K. This crystal is less resistive than D5(3) and when connected to $R_L$=101 kΩ the transition to the mixed state could not be reached. (The measurements for T=305 K were unintentionally not extended to the highest currents, but this had little effect on the conclusions). The traces of |dV/dI| at the bottom of Fig. 1S(b) indicate that |dV/dI|$_{max}$ lie below the 10 kΩ for T=289, 305 and 320 K. However, the behavior of I(V) for 289 K and $R_L$=11 kΩ indicates unstable conditions. Upon switching, the corresponding characteristic cuts the "well behaved" trace for $T_0$= 295 K. The jump occurs along a line of slope of ~ 1/$R_L$. The onset of the mixed state at 295, 305 and 320 K, the reversible state and the hystereses best seen on the traces for P(I) in Fig. 1S(e) are in strong contrast to the corresponding trace for 289 K. For 320 K, P(I) in the reversible regime is linear with dP/dI=1.25 V; the corresponding dp/dj=8.3 V/cm is close to that found for D3(11) with $R_L$=101 kΩ. At 320 K, the ranges of reversible current and resistance are very wide, spread over more than one order of magnitude in the case of R(V). The traces for $R_L$=1 kΩ were obtained under unstable conditions and resemble those in Fig. 2 ((g)-(i)).

$P_{max}(T_0)$ obtained for D3(11) (see Fig. 1S (f) and 1S(i)) is displayed in Fig. 3(b) of the main text.

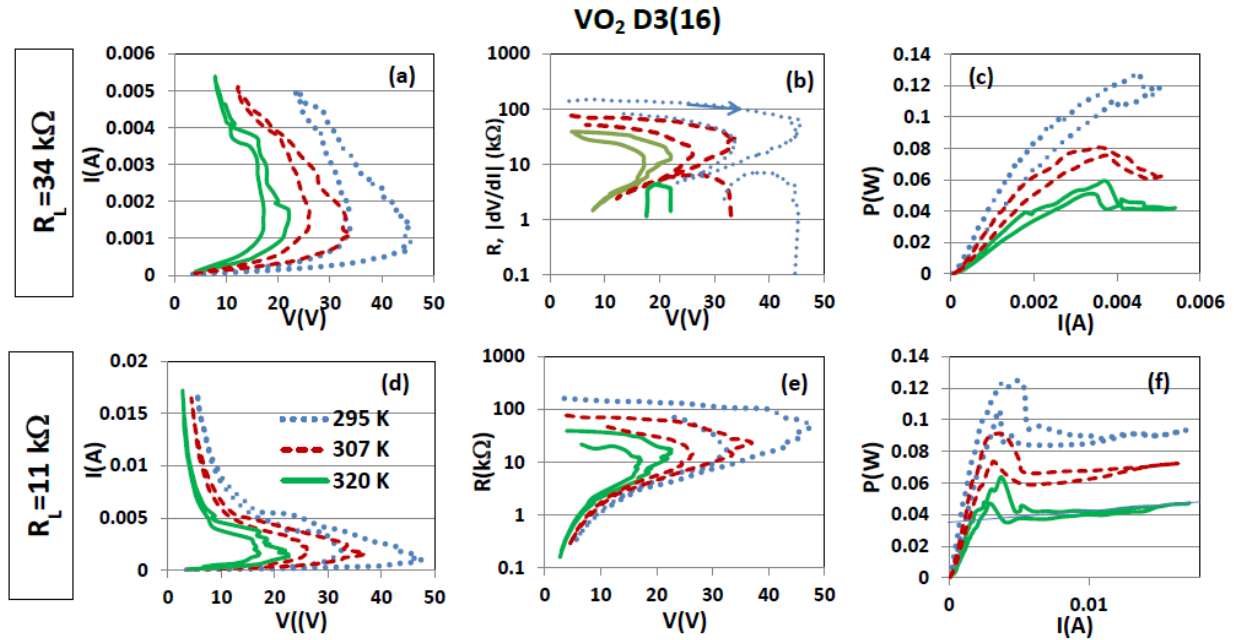

**Figure S4.** I(V), R(V)(=V/I) and P(V)(=VI) for sample D3(16) at temperatures 295, 307 and 320 K (see legend on frame (d)) with $R_L$=34 kΩ ((a) –(c)) and $R_L$=11kΩ ((d) –(f)). |dV/dI|(V) in the NDR regime is shown in (b) below R(V).

The I-V characteristics, R(V) and P(I) for sample D3(16) connected to $R_L=34$ k$\Omega$ are shown in Figs. S4 ((a)-(c)). The traces of $|dV/dI|(V)$ for this $R_L$ are shown below the R(V) traces in Fig. S4(b). They show that for all three temperatures $|dV/dI|_{max}< 10$ k$\Omega$. The respective traces for $R_L=11$ k$\Omega$ are shown in Figs. S4 ((d)-(f)). By decreasing $R_L$ the range of currents was increased by a factor of three and that of the resistance modulation by one order of magnitude.

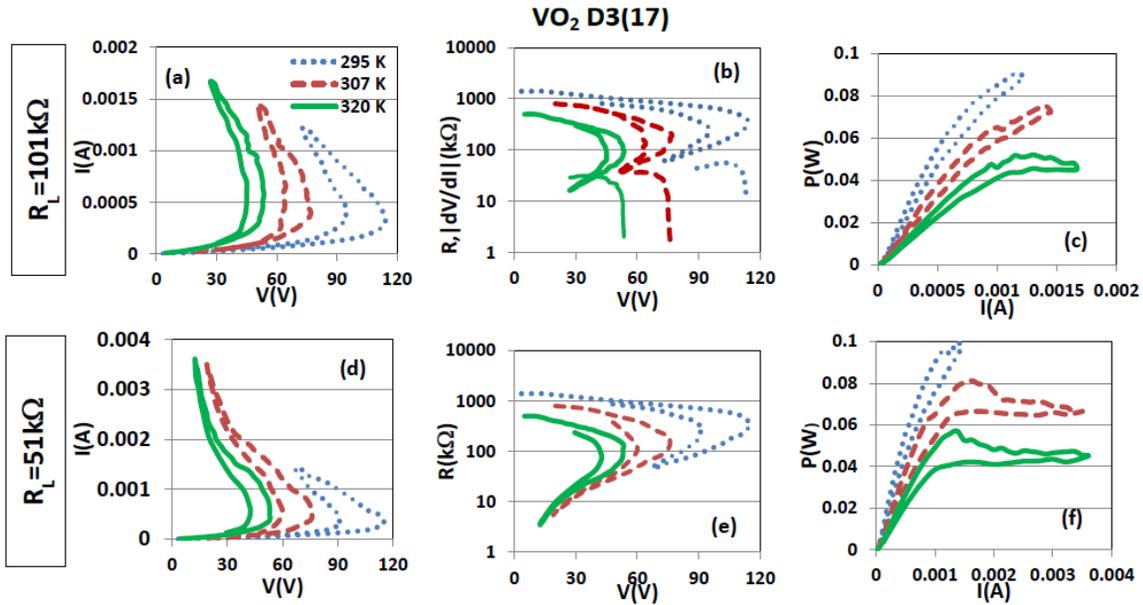

**Figure S5.** I(V), R(V)(=V/I) and P(V)(=VI) for sample D3(17) at temperatures 295, 307 and 320 K (see legend on frame (a)) with $R_L=101$ k$\Omega$ ((a) –(c)) and $R_L=51$k$\Omega$ ((d) –(f)). $|dV/dI|(V)$ in the NDR regime is shown in (b) below R(V).

Sample D3(17) was very thin (see Table I) and of high resistance and consequently $|dV/dI|_{max}$ was high at all three temperatures (see Fig. S5). $R_L$ could not be reduced below 51 k$\Omega$ to maintain steady state conditions. A wide range of reversible P(I) such as those seen in Figs. 4(f) and 2S(f) could not be reached. The traces obtained for this sample emphasize the large reduction of $V_{max}$ with increasing temperature.

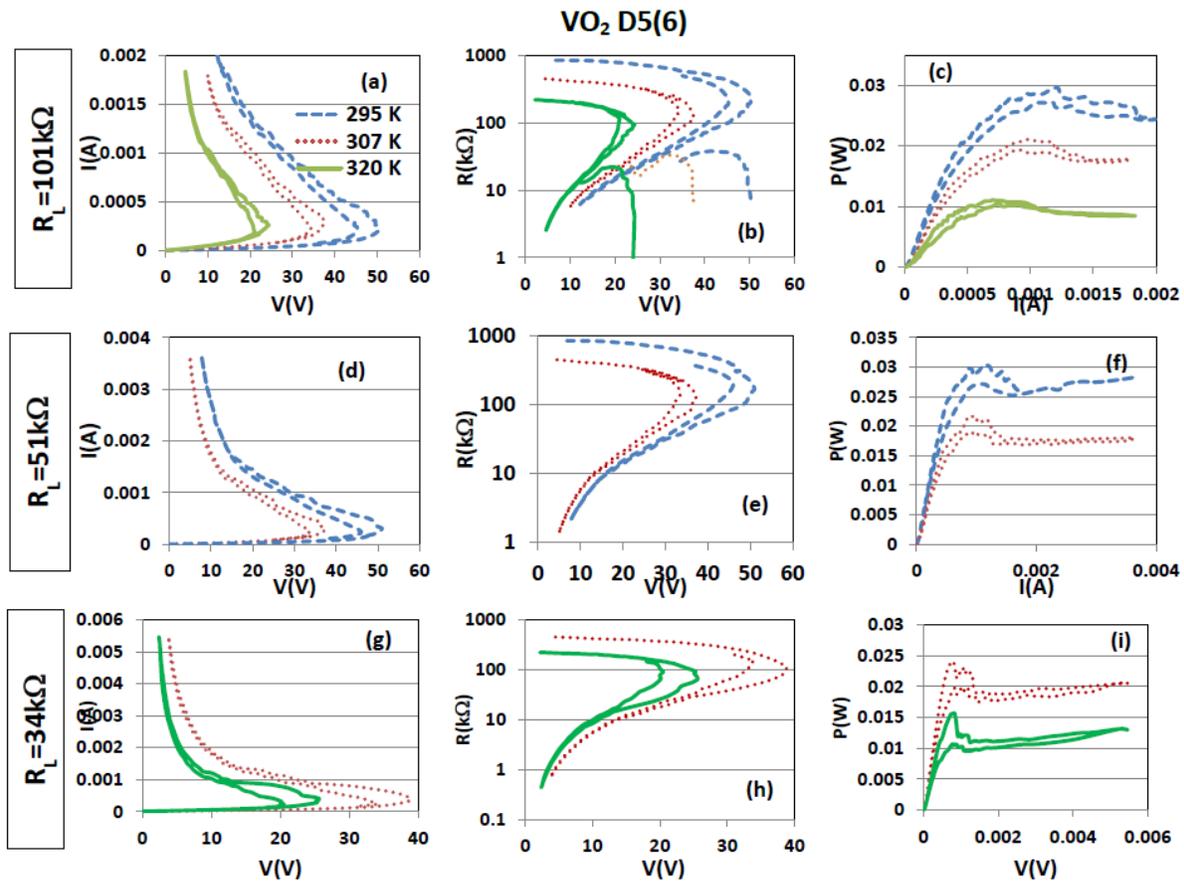

**Figure S6.** I(V), R(V)(=V/I) and P(V)(=VI) for sample D5(6) at temperatures 295, 307 and 320 K (see legend on frame (a) with $R_L <|dV/dI|_{max}$ at all temperatures.

The I-V characteristics, R(V) and P(I) for sample D5(6) connected to $R_L$=101, 51 and 34 kΩ are shown in Figs. S6 ((a)-(i)), each for the temperatures at which the I-V characteristics are at steady state.

$V_{max}$ versus $T_0^{1/2}$ and $|dV/dI|_{max}$ versus $R_0$ obtained for the four samples D3(11), D3(16), D3(17) and D5(6) (see Figs. S3 –S6) were added to Figs. 5(a) and 5(b) of the main text.